\documentclass[%
 reprint,
 superscriptaddress,
 amsmath,amssymb,
 aps,
]{revtex4-2}

\usepackage{graphicx}
\usepackage{dcolumn}
\usepackage{bm}

\begin{document}

\preprint{APS/123-QED}

\title{A parametrically programmable delay line for microwave photons}

\author{Takuma Makihara}
 \email{makihara@stanford.edu}
\author{Nathan Lee}
\affiliation{
 Department of Applied Physics, Stanford University, Stanford, California 94305, USA}
\author{Yudan Guo}
\affiliation{
 Department of Applied Physics, Stanford University, Stanford, California 94305, USA}
\author{Wenyan Guan}
\affiliation{
 Department of Applied Physics, Stanford University, Stanford, California 94305, USA}
\author{Amir Safavi-Naeini}
 \email{safavi@stanford.edu}
 \affiliation{
 Department of Applied Physics, Stanford University, Stanford, California 94305, USA}


\date{\today}

\begin{abstract}
Delay lines that store quantum information are crucial for advancing quantum repeaters and hardware efficient quantum computers. Traditionally, they are realized as extended systems that support wave propagation but provide limited control over the propagating fields. Here, we introduce a parametrically addressed delay line for microwave photons that provides a high level of control over the stored pulses. By parametrically driving a three-wave mixing circuit element that is weakly hybridized with an ensemble of resonators, we engineer a spectral response that simulates that of a physical delay line, while providing fast control over the delay line’s properties. We demonstrate this novel degree of control by choosing which photon echo to emit, translating pulses in time, and even swapping two pulses, all with pulse energies on the order of a single photon. We also measure the noise added from our parametric interactions and find it is much less than one photon. 

\end{abstract}

\maketitle


\section*{Introduction}
Delay lines that preserve quantum information have been proposed as a resource for universal fault-tolerant quantum computing~\cite{pichler2017universal,wan2021fault}. These works propose hardware-efficient approaches to quantum computing where the emission from a single well-controlled qubit is captured and stored in a long delay line to be interacted with at a later time. At optical frequencies, fibers have been used in experiments for generating two-dimensional cluster states, which are a resource for universal quantum computation~\cite{asavanant2019generation,larsen2019deterministic}. In these implementations continuous variable quantum states are generated by squeezing the light field. An important parameter in these systems is inverse of the squeezing bandwidth which approximately determines the temporal extent of a mode and therefore the amount of delay needed to store it. Temporal widths smaller than $10^{-12}~\text{s}$ have been realized in recent experiments~\cite{kashiwazaki2021fabrication} allowing tens of meters of fiber to store on the order of $10^4$ modes simultaneously. In contrast, the quantum emitters used in discrete variable quantum systems typically emit photons much more slowly and need correspondingly longer delays. At microwave frequencies, artificial atoms constructed from superconducting circuits have already been used to demonstrate quantum advantage for certain problems~\cite{arute2019quantum,wu2021strong,jurcevic2021demonstration}. The timescale for emission of a microwave photon from these circuits is around $10^{-6}~\text{s}$. This makes implementing a delay line challenging. For example, a superconducting coplanar waveguide (CPW) on silicon would need to be~$\simeq120$ meters in length to provide enough delay to store a just single mode.

Several innovative approaches have been developed to circumvent traditional delay lines' substantial physical length requirements. The first of these employs slow-wave or slow-light structures, which effectively reduce the speed of electromagnetic wave propagation by using metamaterials~\cite{ferreira2021collapse} or resonator arrays~\cite{bao2021demand,yariv1999coupled}. This technique allows for a significant decrease in the physical size of the delay line while maintaining its function. The second approach involves using waves that are not electromagnetic. A classic example is acoustic waves, such as those used by early mercury delay lines~\cite{auerbach1949mercury}, and more recent Surface Acoustic Wave (SAW) technologies~\cite{maines1976surface} which have been applied for delaying quantum information~\cite{bienfait2019phonon}. These methods use waves with inherently slower propagation speeds, in this case acoustic, to achieve the desired delay in a reasonable amount of space. Lastly, approaches that use atoms or emitters based on Electromagnetically Induced Transparency (EIT)~\cite{eisaman2005electromagnetically}, or those using atomic frequency combs (AFCs)~\cite{afzelius2009multimode,afzelius2010demonstration} have been developed in the last decades with a view towards quantum information. The latter AFC schemes are noteworthy as they emulate the response of a traditional delay line through light-matter interactions facilitated by pump fields. They realize what is in effect a virtual delay line -- even if the system behaves as if a localized pulse is propagating down a waveguide, in reality the excitation is in the coherences of a large number of atoms and may be distributed in space and frequency in a manner that does not resemble a localized propagating pulse. The unique advantage here lies in the ability to alter the characteristics of the delay line simply by modifying the pump fields, offering a dynamic solution for implementing a more robust and versatile delay line.

In this study, we introduce a Parametrically Addressed Delay Line (PADL) as a versatile virtual delay line for microwave photons, leveraging pump fields that drive parametric processes to dynamically control the speed, direction, coupling strength, and connection points of the signal within the line. We implement this virtual delay line by parametrically distributing a data pulse that is launched at a lumped element readout mode into excitations in an ensemble of long-lived resonators. Controlling all of the delay line's properties translates to controlling the parametric drive frequencies, amplitudes, and phases. We show how a parametric delay line gives us more control than a physical waveguide by: (1) controlling the drive amplitudes to selectively choose the number of round trips that a wavepacket makes inside the delay line, (2) controlling the drive phases to translate the pulse in time, and (3) controlling the drive detunings to swap two wavepackets in time. A key question is how the delay line will perform for quantum pulses, and whether the PADL's parametric nature leads to excess noise, such as through parasitic processes, to degrade performance. For this, we measure the added noise from the parametric drives by calibrating the gain of our measurement apparatus using our resonators as quantum microwave parametric oscillators (MPOs)~\cite{wang2019quantum} that operate as quantum-calibrated sources of microwave radiation. Specifically, we use the number of photons in our quantum MPO near threshold as an in-situ noise power calibration device~\cite{berdou2023one}, and find that the added noise is much less than one photon per mode.

\begin{figure*}
\includegraphics[scale=1.0]{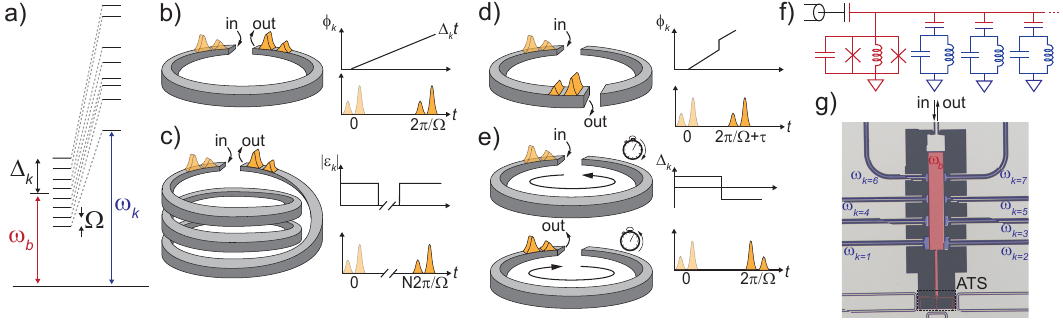}
\caption{\label{fig:PADL_cartoons}Principles of the PADL and its implementation using an ATS. (a) We implement a parametric delay line by parametrically coupling an ensemble of resonators (with frequencies $\omega_k$) to a readout mode (with frequency $\omega_b$, which we refer to as the buffer mode). The parametric drives are indicated by dashed lines. We engineer the buffer mode's spectrum into looking like a delay line with an FSR denoted by $\Omega$ by choosing the converted storage photons to have detunings $\Delta_k$ relative to $\omega_b$ and to form a frequency comb with an FSR given by $\Omega$. (b) Analogy between a physical waveguide delay line (illustrated as a ring supporting pulses) and the PADL under continuous parametric driving. In this system, the \( k^\text{th} \) delay line mode evolves by accruing a phase $\phi_k=\Delta_k t$. The input pulses are illustrated by dashed lines and the delayed pulses are illustrated by solid lines. (c) By turning off all parametric drive amplitudes $|\epsilon_k|$ when a photon echo is about to rephase, one can prevent the signal from emitting into the environment and thereby selectively emit a later photon echo. (d) By instantaneously translating the \( k^\text{th} \) delay line mode's phase by $\phi_k \rightarrow \phi_k + \Delta_k \tau$, one can translate the pulse by a time $\tau$ (modulo $2\pi/\Omega$). (e) By swapping the detunings of the delay line modes $\Delta_k \rightarrow -\Delta_k$ one effectively take $t \rightarrow -t$ in the phase accrued by the delay line modes and swap the order of two pulses.  (f) Circuit diagram of our parametric delay line implementation. We weakly hybridized CPW resonators (illustrated by lumped LC resonators) with one ATS. (g) False-color optical micrograph of our device, where the CPW resonators are shown in blue and the lumped element buffer mode is shown in red. The ATS is indicated by a dashed box.}
\end{figure*}

\section*{\label{sec:level1}Results}
\subsection*{\label{sec:level1}Implementing a parametric delay line}
The PADL works by parametrically distributing a data pulse that is launched at a readout mode (referred to as the buffer mode) into excitations of a collection of long-lived storage resonators. By controlling the parametric drives, we can engineer the buffer mode's spectrum to emulate the characteristics of a reflective delay line. As a simple example, consider using PADL to emulate a physical delay line with a free spectral range (FSR) given by $\Omega$. This can be achieved with PADL by continuously parametrically coupling the buffer mode and the storage resonators such that the detuning $\Delta_k$ between the converted photons from the \( k^\text{th} \) storage resonator and the buffer form a frequency comb with an FSR (peak spacing) $\Omega$ and such that the coupling strengths realize the desired loading. Fig.~1a illustrates the relevant frequencies in this example. The storage resonator frequencies are labeled by $\omega_k$ and the buffer frequency is labeled by $\omega_b$. The parametric couplings are illustrated by dashed lines and the parametrically converted storage photons are shown to have detunings $\Delta_k$ with an FSR given by $\Omega$. The storage resonator frequencies do not need to be precisely placed or tuned -- we use the parametric drive frequencies to compensate for any disorder to realize the mode spacing $\Omega$~\cite{lee2020propagation}. Fig.~1b illustrates this mode of operation by introducing an analogy to a physical delay line (illustrated as a ring supporting data pulses). For continuous parametric coupling, the delay line modes evolve by accruing a phase $\phi_k = \Delta_k t$ and rephase after a round-trip time $T_\text{rt} = 2\pi/\Omega$, such that the output pulse (drawn with a solid line) is simply the input pulse (drawn with a dashed line) delayed by $T_\text{rt}$.

However, the PADL provides far more exotic dynamics than a simple physical delay because we can parametrically program the delay line mode couplings, phases, and detunings. Crucially, these couplings are independently tunable; changing the \( k^\text{th} \) parametric drive changes the coupling to the \( k^\text{th} \) storage resonator. In Fig.~1c, we illustrate how we can effectively disconnect the virtual delay line from the input/output by shutting off the parametric drives, i.e. by setting their amplitudes $|\epsilon_k| = 0$. This prevents the rephased signal from being emitted into the environment and causes it to propagate around the virtual delay line for longer. Turning the drives back on reconnects the waveguide to the environment and causes the pulse to be re-emitted at the next round-trip time $N T_\text{rt}$ for some positive integer $N$. This mode of operation is analogous to fiber loop buffers which have been recently considered for storing quantum information~\cite{lee2023fiber}. Full control over the phases of the drives allows us arbitrary access to information stored in the delay line. In Fig.~1d, we illustrate how we continuously translate the pulse forwards or backwards in time by an amount $\tau$ (modulo $T_\mathrm{rt}$) by instantaneously translating each phase by $\phi_k \rightarrow \phi_k + \Delta_k \tau$. This is analogous to accessing the pulse at positions other than the open port in the virtual delay line. In Fig.~1e, we illustrate how controlling the parametric drive frequencies allows us to swap two pulses in time. Specifically, changing $\Delta_k \rightarrow -\Delta_k$ is equivalent to taking $t \rightarrow -t$ in terms of the phase that the virtual delay line modes accrue. This is analogous to switching the pulse propagation direction in the virtual delay line. While we focus on these four experiments, one can engineer more complicated pulse dynamics with the novel degree of control provided by a parametric delay line. 

We implement the physical circuit on-chip by fabricating quarter-wavelength CPW resonators that are weakly capacitively coupled with a nonlinear resonant circuit known as an Asymmetrically Threaded SQUID (ATS)~\cite{lescanne2020exponential}. The ATS circuit is given by two nominally identical Josephson junctions forming a loop with an inductive shunt in the center of the loop. We use an array of Josephson junctions to form the inductive shunt. The circuit diagram is illustrated in Fig.~1f, where we illustrate the CPW resonators by their equivalent lumped element LC models. A false-color optical micrograph of the PADL device is shown in Fig.~1g, where the buffer mode is highlighted in red and the CPW resonators are highlighted in blue.

The ATS provides the lumped element buffer mode that we use to readout our pulses, as well as the necessary nonlinearity to parametrically couple the different modes of our circuit. We choose to use an ATS for three reasons. Firstly, the ATS provides three-wave mixing as opposed to four-wave mixing -- the native nonlinearity of Josephson junctions. Three-wave mixing allows us to parametrically swap between the ATS lumped element mode and the CPW modes while minimizing the adverse effects from four-wave-mixing interactions that are characteristic of a multimode systems connected to junctions. Secondly, the ATS has an inductor, which provides an unconfined parabolic potential and therefore its lumped element mode can be strongly driven before it becomes nonlinear. Finally, we can use the three-wave mixing nonlinearity to implement a quantum MPO in the CPWs, which we can use for noise power calibration. 

When the ATS is precisely biased at its ``saddle-point,'' the energy associated with the phase drop across the junctions changes from the usual $\cos(\varphi)$ to $\sin( \varphi)$. This crucial modification shifts the dominant nonlinear term from $\varphi^4$ to $\varphi^3$. We use this altered nonlinearity to enable the three-wave mixing processes essential for the operations conducted here. The Hamiltonian at the saddle-point is
\begin{equation}
\begin{split}
    \hat{H} & = \hbar\omega_b \hat{b}^\dagger \hat{b} + \sum_k \hbar\omega_k \hat{a}_k^\dagger \hat{a}_k \\
    & -2E_J \epsilon_p(t)\sin\Big(\varphi_b(\hat{b}+\hat{b}^\dagger) + \sum_k \varphi_k (\hat{a}_k+\hat{a}_k^\dagger) \Big),
\label{eq:Full_Hamiltonian}
\end{split}
\end{equation}
where $\varphi_b$ is the node flux zero-point fluctuation (ZPF) of the buffer mode across the ATS Josephson junctions with annihilation operator $\hat{b}$ and frequency $\omega_b$. Similarly $\varphi_k$ is the ZPF of the \( k^\text{th} \) CPW resonator fundamental mode at the same circuit node with annihilation operator $\hat{a}_k$ and frequency $\omega_k$. $E_J$ is the individual junction energy in the SQUID, and $\epsilon_p(t)$ is a time-dependent parametric flux pump threading the SQUID. Note that we have assumed $|\epsilon_p(t)|\ll 1$.

\begin{figure}[b]
\includegraphics[scale=1.0]{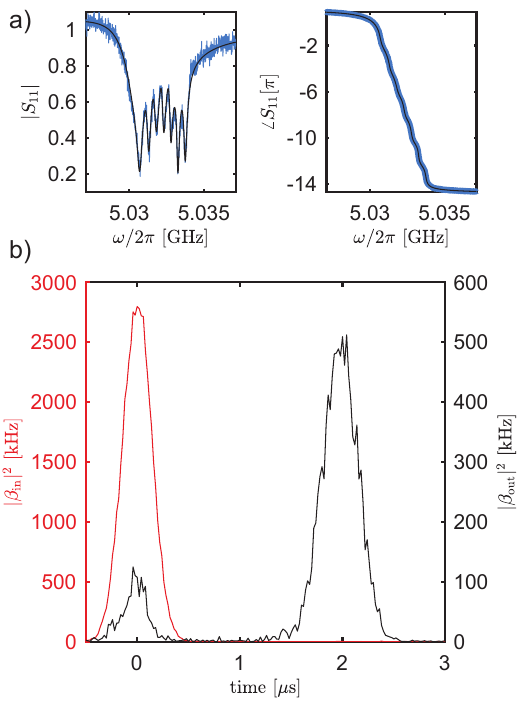}
\caption{\label{fig:S_params} A continuously coupled parametric delay line. (a) The reflection coefficient of our buffer mode ($S_{11}(\omega)$) when seven CPW resonators are parametrically coupled to the mode. (b) Analog-to-digital converter traces of a pulse that is stored (black) and not stored (red) in our parametric delay line. The delayed pulse is delayed by approximately $T_\text{rt} = 2~\mu$s $\simeq 2\pi/\Omega$. The time-domain traces are reported in units of photon flux (see Supplementary Note 6 for more details).}
\end{figure}

We resonantly select specific interactions by driving the buffer mode while simultaneously flux pumping the SQUID. We focus here on a beamsplitter interaction between the buffer mode and the \( k^\text{th} \) CPW's fundamental mode. We flux pump the SQUID at a single frequency $\omega_p$ and drive the buffer mode with multiple drives tones, as captured by the following driving Hamiltonian:
\begin{equation}
\hat{H}_\mathrm{drive}/\hbar  = \sum_k \big( \epsilon_k \hat{b} e^{i\omega_{d,k}t} + \mathrm{h.c.} \big).
\label{eq:Driving_Hamiltonian}
\end{equation}
Here, $\omega_{d,k}$ is the frequency of a detuned drive on the buffer with field amplitude $|\epsilon_k|$. We choose the drive frequencies to satisfy
\begin{equation}
\omega_{d,k} = \omega_p - (\omega_b + \Delta_k) + \omega_k.
\label{eq:Drive_Frequency}
\end{equation}
In the frame where both the CPW mode and the buffer are rotated out, our total Hamiltonian approximately becomes~\cite{chamberland2022building}:
\begin{equation}
\hat{H}/\hbar = \sum_k \Delta_k \hat{a}_k^\dagger \hat{a}_k + \sum_k g_k \hat{a}_k^\dagger \hat{b} +  \mathrm{h.c.}
\label{eq:Multimode_BS_Hamiltonian}
\end{equation}
\noindent where $\hbar g_k = E_J \epsilon_p \varphi_b^2\varphi_a \beta_k$ is the parametric coupling strength between the buffer and the \( k^\text{th} \) CPW's fundamental mode, and $\beta_k$ is the small displacement on $\hat{b}$ generated by the \( k^\text{th} \) drive. By operating the buffer in the fast-cavity regime, i.e., $\kappa_{b,e} \gg g_k$, we can adiabatically eliminate it ($\kappa_{b,e}$ is the extrinsic loss rate of the buffer mode). We tune the drive amplitudes and frequencies so that the $g_k$ and harmonically placed $\Delta_k$ in the resulting effective Hamiltonian closely resemble that of a delay line. Importantly, by controlling the parametric drives' frequency, amplitude, and phase, we control the corresponding delay line mode’s detuning, coupling, and phase. In principle, the flux pump frequency is arbitrary because the drive frequencies can be chosen to satisfy Eq.~\ref{eq:Drive_Frequency}. In practice, one could be limited by the amount of microwave power that is available to drive the buffer.

\begin{figure}[b]
\includegraphics[scale=1.0]{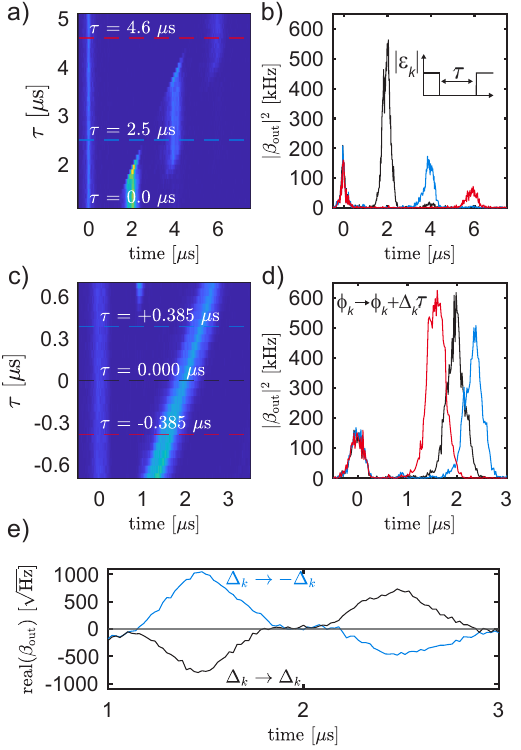}
\caption{\label{fig:pulsed_experiments} Parametric control of the delayed pulse. (a,b) By turning off the parametric drive amplitudes for a duration $\tau$, one can selectively emit later photon echos. (a) A color map of ADC traces for $\tau$ swept from $0~\mu s$ to $5~\mu s$. (b) ADC traces for $\tau$ = $0.0~\mu s$ (black), $2.5~\mu s$ (blue) and $4.6~\mu s$ (red). (c,d) By instantaneously translating the phase of the \( k^\text{th} \) parametric drive by an amount $\Delta_k \tau$, one can translate the pulse in time by $\tau$ (modulo $2\pi/\Omega$). (c) A color map of ADC traces for $\tau$ swept from $-0.7~\mu s$ to $+0.7~\mu s$. (d) ADC traces for $\tau$ = $-0.385~\mu s$ (red), $0.000~\mu s$ (black), and $+0.385~\mu s$ (blue). (e) By instantaneously swapping the detunings $\Delta_k \rightarrow -\Delta_k$ in the parametric delay line, one can swap two pulses in time.}
\end{figure}

The ATS parameters are chosen such that $\omega_b/2\pi=5.0073$~GHz and $\omega_k/2\pi \simeq~6.91-7.46$~GHz (see Supplementary Note 1 for more details). The buffer is capacitively coupled to the environment at a rate $\kappa_{b,e}/2\pi=3.95$~MHz, and the resonator intrinsic quality factors approximately range from $110\times10^3$ to $300\times10^3$. We estimate from finite element electromagnetic simulations (see Supplementary Note 1 for more details) that the buffer impedance and hybridization strengths with the CPWs leads to flux ZPF across the junction for each mode of roughly $\varphi_b=0.336$ and $\varphi_k \simeq0.018-0.023$.

\subsection*{\label{sec:level1}Parametric control of stored wavepackets}
We first start with the simplest PADL experiment -- implementing a response that mimics that of a reflectively terminated transmission line probed at the other end. For this, we tune and fix the parametric drives' amplitudes, phases, and detunings (as illustrated in Fig.~1b) for seven of the CPWs. For all the experiments in this work, we only parametrically couple seven of the eight CPW resonators to the buffer as we observed one of the CPW modes to have larger frequency fluctuations, which we attribute to a nearby two-level system (TLS) defect. We set the FSR to be $\Omega/2\pi = 500~\mathrm{kHz}$ so the delay line bandwidth closely matches the buffer mode's extrinsic coupling rate. The continuous wave (CW) flux pump tone is provided by a signal generator (Keysight E8257D PSG). The parametric drive intermediate frequency (IF) tones are all generated on a single arbitrary waveform generator (AWG) channel (Tektronix 5200) before being up-converted to drive the buffer mode. Before being up-converted, the AWG output is amplified by a room-temperature low-noise amplifier. All pulses sent into the delay line are played and demodulated using an Operator-X (OPX) from Quantum Machines Inc. (QM), and similarly the pulses are up-converted and down-converted using an Octave from QM. The demodulated pulses are also digitized on an analog-to-digital converter (ADC) in the OPX (see Supplementary Note 2 for more details). We emphasize that the flux pump tone provides the magnetic flux $\epsilon_p(t)$ through the ATS loops and is fed to the device via a transmission line that is grounded near the ATS, whereas the parametric drives $\epsilon_k(t)$ are fed through the readout transmission line that is capacitively coupled to the buffer mode (see Supplementary Note 2 for more details).

We probe the PADL on reflection near the buffer mode frequency and clearly observe the parametrically coupled modes in the normalized reflection coefficient $S_{11}(\omega)$ (Fig.~2a). We tune the parameters to obtain an approximately linear phase response centered at $5.0321$~GHz to emulate the $\omega T_\text{rt}$ phase response of a physical delay line. We measure the time-domain response by sending pulses at the PADL. Fig.~2b shows the results of reflecting an attenuated Gaussian pulse ($\langle n \rangle \simeq 1$, with a temporal FWHM of 471 ns; see Supplementary Note 6 for more details on the calibration of the attenuation). The red pulse centered at $t = 0$ results from reflecting a pulse off of the device in the absence of pump and drives and with the  buffer mode far-detuned, so that the device acts as a mirror. The black pulse results from reflecting a pulse off the buffer mode while the device is emulating a delay line. We observe that the pulse is approximately delayed by $2~\mu\text{s}\simeq T_\mathrm{rt}$. The small reflected pulse at $t=0$ in the black trace is due to mismatched impedance between the environment and the parametric delay line, including any inevitable mismatches from device packaging (see Supplementary Note 5 for more details). The time-domain traces are reported in units of input photon flux $|\beta_\text{in}|^2$ and output photon flux $|\beta_\text{out}|^2$. These fluxes are related by the input-output boundary condition $\beta_\mathrm{out} = \beta_\mathrm{in} + \sqrt{\kappa_{b,e}}\beta$, where $\beta = \langle \hat{b} \rangle$.

By turning the parametric drive amplitudes $|\epsilon_k|$ off, we prevent a stored wavepacket from being emitted, causing the pulse to undertake another round trip in the delay line, as illustrated in Fig~1c. Turning the amplitudes back on allows the next photon echo to be emitted. In Fig.~3a, we sweep the duration $\tau$ over which we turn off the parametric drive amplitude. Delay $\tau=0$ corresponds to the drives being on continuously. We see that as $\tau$ exceeds $T_\text{rt}$, the first echo disappears and the second echo at $2 T_\text{rt}$ becomes more prominent. Similarly, as $\tau$ exceeds $2 T_\text{rt}$, the second photon echo disappears and the third photon echo at $3 T_\text{rt}$ becomes more prominent. In Fig.~3b, we show the ADC data of traces taken with $\tau$ = $0~\mu$s (black), $2.5~\mu$s (blue), and $4.6~\mu$s (red). 

We fine-tune the photon emission time by modifying the parametric drive phases to modify $\phi_k$. By changing these phases we continuously translate the pulse in time, as illustrated in Fig.~1d. Specifically, by translating $\phi_k \rightarrow \phi_k + \Delta_k \tau$, we can translate our pulse by a time $\tau$ modulo $T_\text{rt}$. In Fig.~3c, we sweep the duration $\tau$ by which we translate our pulse, where $\tau=0$ corresponds continuous parametric driving without phase modification. We see that as we sweep $\tau$, the time at which the pulse re-emits translates linearly. In Fig.~3d, we show the ADC data of traces taken with $\tau$ = $0.000~\mu$s (black), $-0.385~\mu$s (red), and $+0.385~\mu$s (blue).

Finally, we show how controlling the detunings of the parametrically converted CPW photons $\Delta_k$ allows us to swap two pulses in time, as illustrated in Fig.~1e. Swapping the detunings $\Delta_k \rightarrow -\Delta_k$, is analogous to performing time reversal $t \rightarrow -t$ in the phase accrued by the delay line modes, causing the first stored pulse to be emitted after the second pulse. In Fig.~3e, we show two complex ADC traces: one where the detunings are swapped (blue) and one where they are not (black). In this experiment, we send two pulses with FWHM = $283$~ns and separation of $1000$~ns. We make the first pulse have a $\pi$ phase shift relative to the second pulse and plot one quadrature of the measured ADC data to clearly show that the two pulses are swapped in order when the detunings are swapped. The black trace in Fig.~3e plots the case where the detunings are not swapped, corresponding to the case of continuous drives. The blue trace in Fig.~3e plots the case where the detunings are swapped, corresponding to the case where we take $t \rightarrow -t$. We clearly see that relative to the black trace, the pulses have been swapped. In Supplementary Note 5, we provide additional numerical simulations considering two pulses with different relative amplitudes and phases.

\begin{figure}[b]
\includegraphics[scale=1]{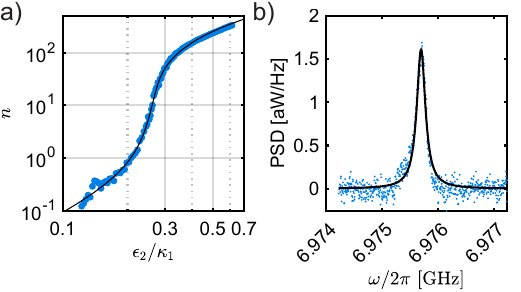}
\caption{\label{fig:OPO_cal} Calibration of added noise. (a) A plot of the number photons $n$ vs. normalized two-photon drive strength $\epsilon_2/\kappa_1$ as one MPO crosses threshold. Both axes are log scale. (b) Spectrum of the same resonator when the pump and drives used to generate a parametric delay line are applied on the device.}
\end{figure}

\subsection*{\label{sec:level1}Fidelity and added noise}
An important figure of merit in quantum memories is the fidelity of the state being read out compared to that which was stored. Ideally, the output pulse should be identical in amplitude and phase to the input pulse. In reality, distortions and loss induced by storage in the PADL and thermal noise reduce the fidelity. We will show that the latter is negligible in our system, and first focus on distortion and loss. Given two pulses that are characterized by temporal mode operators $\hat{A}_1 = \int \mathrm{dt} f(t) \hat{a}(t)$ and $\hat{A}_2 = \int \mathrm{dt} g(t) \hat{a}(t)$, we define the fidelity as $F = |\int dt f^*(t)g(t)|^2$, where $f(t)$ is the temporal mode profile for the input pulse and $g(t)$ is the temporal mode profile for the delayed pulse. Here, we focus on the fidelity of the pulse that is delayed in the presence of continuous parametric drives (such as the black pulse in Fig. 2b) such that $g(t)\simeq f(t-T_\text{rt})$. We calculate these mode profiles by normalizing the detected field measured on the ADC. The input mode is always normalized such that $\int dt |f(t)|^2=1$ corresponding to unit detection probability. To isolate the impact of distortions on the pulse while it is stored in the PADL, we first normalize the delayed pulse such that $\int dt |g(t)|^2=1$. We also exclusively focus on the pulse centered at $2~\mu s$ and neglect contributions from the small reflected pulse at $t=0$, which can be caused by impedance mismatches in the device packaging. With this normalization for $g(t)$, which ignores losses and effectively compares the shape of the delayed temporal mode to the input temporal mode, we measure $F = 0.95$. Time-domain simulations of the semi-classical equations of motion for $\hat{a}(t)$ and $\hat{b}(t)$ derived from Eq.~\ref{eq:Multimode_BS_Hamiltonian} show that fidelity can be further improved by optimizing the parametric delay line parameters and input pulse bandwidth (see Supplementary Note 5 more details). To include the effect of losses, we normalize $g(t)$ by the same factor that we normalize $f(t)$ resulting in $\int dt |g(t)|^2\le1$. In this case, we find a $F = 0.21$ that includes both loss and distortion, showing that the infidelity is primarily due to losses in the storage resonators. The infidelity from storage resonator loss is neither limited by the loss of the worst resonator nor the sum of all resonator losses. Given that the input pulse is parametrically distributed into the collection of storage resonators, the input pulse's frequency components near $\Delta_k$ are predominantly attenuated by the \(k^\text{th}\) resonator's loss rate.

In addition to loss and distortion, an important concern is whether the parametric driving of the PADL leads to excess noise being added to the microwave field. We estimate the added noise by measuring the power spectral density of the field emitted at the CPW mode frequencies when the drives are on. The key challenge is to calibrate the gain and loss in the readout signal path with sufficient precision to be able to infer the microwave fluctuations at the device from the field detected outside the fridge. Previously, this has been accomplished by using in-situ calibrated sources, such as shot noise tunnel junctions~\cite{spietz2003primary} or qubits~\cite{macklin2015near}. In both these cases, the physics of the source is sufficiently well-understood to provide a signal with a well-defined photon flux or noise power without requiring a component-by-component accounting of gain and loss. Here we use the device itself, operated as a parametric oscillator, as such a quantum calibrated source. In a classical model of a parametric oscillator there is a discontinuity at the oscillation threshold. This discontinuity is smoothed away by the fluctuations of the electromagnetic field in a more accurate quantum model of a parametric oscillator. The dependence of number of intracavity photons $n$ vs. drive power obtains a characteristic shape~\cite{wang2015coherent,berdou2023one} which we use to infer the number of photons and calibrate the gain of our amplification chain (see Supplementary Note 6 for more details).  

We implement the MPO on the same device by pumping the ATS at $\omega_p = 2\omega_k-\omega_b$ to resonantly select a 2-photon swapping term given by $\hat{a}_k^2\hat{b}^\dagger +\hat{a}_k^{2\dagger}\hat{b}$. After adiabatically eliminating the fast-decaying buffer mode and applying a resonant drive, the effective dynamics of the $k^\text{th}$ CPW mode are governed by a Lindblad master equation $\dot{\hat \rho} = -i[\hat H,\hat \rho]+\mathcal D[\hat L_1]\hat \rho+\mathcal D[\hat L_2]\hat \rho$, with the Hamiltonian and loss operators given by 
\begin{equation}
    \hat{H}  = i\epsilon_2 \hat{a}^2 + \mathrm{h.c.},~~
    \hat{L}_2  = \sqrt{\kappa_2} \hat{a}^2,~~\text{and}~~ 
    \hat{L}_1  = \sqrt{\kappa_1} \hat{a}.
\label{eqn:mpo_eqn}
\end{equation}
In these equations, $\epsilon_2$ is the two-photon drive strength, $\kappa_2$ is the two-photon loss rate, and $\kappa_1$ is the single-photon loss rate. 

In Fig.~4a, we show how the steady-state CPW intracavity photon number $n$ changes with the amplitude of the normalized two-photon drive when we operate one of the CPWs (the resonator at 6.975562 GHz) as a quantum MPO. In the following experiments, we directly readout the individual CPWs through the readout port, which is possible due to weak parasitic capacitances. We measure the integrated power spectral density (PSD), which is proportional to $n$, on an RF spectrum analyzer and plot the result versus the amplitude of the driving field sent to the buffer mode. A smooth transition is clearly visible and the data agrees closely with the quantum model (solid black line) of the MPO (see Eq.~\ref{eqn:mpo_eqn}) with three fitting parameters: (1) the proportionality constant between $\epsilon_2/\kappa_1$ and the driving field at the instrument, (2) the proportionality constant between $n$ and the integrated PSD at the instrument, which is related to the gain, and (3) $\kappa_2/\kappa_1$. 

With the gain calibrated, we can determine the number of added noise photons. In Fig.~4b, we show the spectrum of the same CPW resonator as in Fig.~4a but now with the parametric drives and pump for the delay line experiments turned on. By fitting this spectrum to a Lorentzian (solid black line) to extract the spectral area and by using our calibrated gain, we conclude that 0.11 noise photons are added to this mode when it is operated as the PADL. We find that the added noise in all modes is always less than $\simeq$ 0.15 photons for the reported drive intensities. The added noise is probably due to heating from our strong parametric flux pump. This pump power is comparable to previous quantum-limited parametric amplifiers~\cite{frattini2018optimizing,sivak2019kerr} and is likely comparable to previous ATS-based devices with higher quality-factor resonators demonstrating cat states~\cite{lescanne2020exponential,reglade2023quantum}. Therefore, we do not believe this small added noise will hinder future quantum applications for the PADL. 

\section*{\label{sec:level1}Discussion}
Unlike a waveguide delay line, the PADL gives us complete control over the detunings, phases, and coupling rates of the delay line modes. Furthermore, delays that are comparable to a several kilometer long waveguide can be achieved for microwave photons in a small footprint. Unlike catch-and-release methods that require a single resonant mode and precise mode matching via dynamic and precisely timed control of cavity parameters, PADL offers far greater flexibility. By using a three-wave mixing circuit element  we sidestep issues due to parasitic processes that arise more frequently in four-wave mixing schemes. In lieu of performing process tomography on encoded qubits~\cite{eichler2011experimental}, which would enable the evaluation of the delay as a quantum process, we characterize the bosonic channel by measuring the number of photons of noise that are added into our delay line as well as the overlap between the input and output wavepackets. We demonstrate dynamic programmability by selecting emission of later photon echos, translating pulses continuously in time, and swapping two pulses stored in the emulated delay line. 

Finally we stress that more intricate control can be engineered, given that the PADL is fully programmable and that integration with qubits is a possibility. Furthermore, by integrating qubits on the same chip as the PADL, one can address any impedance mismatches that arise from packaging. Nonetheless, practical use as a quantum memory will require much higher fidelities. One simple way to improve fidelity is to use larger bandwidth pulses and shorter delays such that the delay is much less than the CPW resonator lifetime. Ultimately, improving the resonator lifetime is important for improving fidelity. In future work, integration with recently developed high-Q Tantalum CPW resonators~\cite{crowley2023disentangling} or microwave cavities~\cite{chakram2021seamless,milul2023superconducting} coupled to ATSs~\cite{lescanne2020exponential,berdou2023one,reglade2023quantum} may open the route to long programmable delays in quantum processors with much higher fidelity. 

\section*{Methods}
\subsection*{Device Fabrication}
Our device is patterned in aluminum on $525~\mu$m thick high-resistivity silicon ($\rho >$ 10 k$\Omega \cdot$cm). The sample is first solvent cleaned in acetone and isopropyl alcohol, followed by the following four-mask process:
\begin{enumerate}
    \item Etched alignment marks: Alignment marks are patterned with photolithography (Heidelberg MLA150 direct-writer) and etched into the sample using XeF$_2$. The sample is then cleaned in baths of piranha (3:1 H$_2$SO$_4$:H$_2$O$_2$) and buffered oxide etchant. 
    \item Circuit patterning: Ground planes, CPWs, flux lines, and the ATS island are patterned with photolithography, followed by a gentle oxygen plasma. Aluminum is deposited in an electron beam evaporator (Plassys) and lifted off in N-Methyl-2-pyrrolidone (NMP). 
    \item Junction patterning: The ATS itself (including the SQUID loop and the superinductor) are patterned by electron beam lithography (Raith Voyager), followed by a gentle oxygen plasma. Aluminum is deposited at an angle of 62$^{\circ}$, followed by oxidation at 50 Torr for 10 minutes, followed by aluminum deposition at an angle of 0$^{\circ}$. Liftoff is performed in NMP. The junction-array inductor is formed from 21 junctions using a Dolan-bridge method, whereas the single junctions in the SQUID are formed using a T-style process~\cite{kelly2015fault}.
    \item Bandaging: We use a bandage mask to ensure a superconducting connection between the previous two masks. The bandages are patterned with electron beam lithography and overlap both masks. Prior to aluminum deposition (at 0$^{\circ}$), we ion-mill in-situ to clear away any oxide, thus ensuring a superconducting connection. Liftoff is performed in NMP.
\end{enumerate}

\subsection*{Circuit Analysis}
The ATS consists of a SQUID (with individual junction energies $E_J$) that is threaded by an inductor (with energy $E_{L_b}$). In terms of the node flux operator $\hat{\varphi}$ of the ATS node, the potential from the inductor and SQUID can be written as:
\begin{equation}
    U = \frac{1}{2}E_{L_b}\hat{\varphi}^2 - 2E_{J}\cos(\varphi_\Sigma)\cos(\hat{\varphi}+\varphi_\Delta)
\end{equation}
where:
\begin{equation}
    \varphi_\Sigma = (\varphi_\mathrm{ext,1} + \varphi_\mathrm{ext,2})/2
\end{equation}
\begin{equation}
    \varphi_\Delta = (\varphi_\mathrm{ext,1} - \varphi_\mathrm{ext,2})/2
\end{equation}
and where $\varphi_\mathrm{ext,1}$ and $\varphi_\mathrm{ext,2}$ are the external magnetic fluxes threading the left and right loops formed by the inductor and a Josephson junction \cite{lescanne2020exponential,berdou2023one,reglade2023quantum}.

When the device is flux-biased to $\varphi_\Sigma = \varphi_\Delta = \pi/2$ and a small RF modulation $\epsilon_p(t)$ is applied to $\varphi_\Sigma$, the potential (to first order in $\epsilon_p$) becomes:
\begin{equation}
    U = \frac{1}{2}E_{L_b}\hat{\varphi}^2 - 2E_{J}\epsilon_p(t)\sin(\hat{\varphi})
\end{equation}

Since the ATS is capacitively coupled to other modes, the node flux operator $\hat{\varphi}$ representing the flux across the junction can be written in terms of the normal modes of the linear circuit as:
\begin{equation}
    \hat{\varphi}~=~\varphi_b(\hat{b}+\hat{b}^\dagger)~+~\sum_k\varphi_k(\hat{a}_k+\hat{a}_k^\dagger)
\end{equation} where $\varphi_b$ is the node flux ZPF of the ``buffer-like" normal mode, and $\varphi_k$ is the node flux ZPF of the ``CPW-like" normal mode of the \( k^\text{th} \) CPW mode \cite{rajabzadeh2023analysis}. The full Hamiltonian is then: 
\begin{equation}
\begin{split}
    \hat{H} & = \hbar\omega_b \hat{b}^\dagger \hat{b} + \sum_k \hbar\omega_k \hat{a}_k^\dagger \hat{a}_k \\
    & -2E_J \epsilon_p(t)\sin\Big(\varphi_b(\hat{b}+\hat{b}^\dagger) + \sum_k \varphi_k (\hat{a}_k+\hat{a}_k^\dagger) \Big)
\end{split}
\end{equation}
as discussed above. 

Microwave Parametric Amplification: When \( \epsilon_p(t) \) is pumped at a frequency \( \omega_p = 2\omega_k - \omega_b \), we need to look for terms in the Hamiltonian that can resonate with this frequency. They are of the form~\cite{chamberland2022building}:
\[ \varphi_b \varphi_k^2 (\hat{a}_k^2 \hat{b}^\dagger + \hat{a}_k^{\dagger 2} \hat{b}). \] 
These terms represent processes where two photons from the \( k^\text{th} \) CPW mode are either absorbed or emitted in combination with the emission or absorption of a photon from the buffer-like mode.

Beamsplitter operation: Alternatively, when \( \epsilon_p(t) \) is pumped at a frequency \( \omega_p = \omega_b + \omega_{d,k} - \omega_k \), and a second drive at frequency \( \omega_{d,k} \) is applied to the buffer, we will implement a beam splitter interaction between the buffer-like mode and the \( k^\text{th} \) CPW mode. In this setup, the interaction Hamiltonian can be simplified to terms that resonate with the combined effect of both drives. The resonant terms in this case are~\cite{chamberland2022building}:
\[ \varphi_b^2 \varphi_k (\hat{b}^\dagger \hat{a}_k  + \mathrm{h.c.}) \]

\section*{Data Availability}
The source data for Fig. 2, Fig. 3, and Fig. 4 generated in this study are provided with the paper and its supplementary information files.

\bibliography{main_bib}

\section*{\label{sec:Acknowledgements} Acknowledgements}
The authors thank O. Hitchcock, R. Gruenke, M. Maksymowych, T. Rajabzadeh, Z. Wang, W. Jiang, F. Mayor, S. Malik, S. Gyger, and E. A. Wollack for useful discussions and assistance with fabrication. The authors thank K. Villegas and T. Gish from QM for their assistance with the Octave and OPX. The authors wish to thank the following sources of financial support for this work: NTT Research, the National Science Foundation CAREER award No.~ECCS-1941826, the Q-NEXT DOE NQI Center, and Amazon Web Services Inc. This material is based upon work supported by the Air Force Office of Scientific Research and the Office of Naval Research under award number FA9550-23-1-0338. Any opinions, findings, and conclusions or recommendations expressed in this material are those of the author(s) and do not necessarily reflect the views of the United States Air Force or the Office of Naval Research. Device fabrication was performed at the Stanford Nano Shared Facilities (SNSF) and the Stanford Nanofabrication Facility (SNF), supported by the NSF award ECCS-2026822. T.M. acknowledges support from the National Science Foundation Graduate Research Fellowship Program (grant no. DGE-1656518).

\section*{Author Contributions}
A.S.N. conceived of the project. T.M. designed, fabricated, and measured the device with assistance from N.L., Y.G., and W.G. The manuscript was written by T.M. and A.S.N with comments from all other authors.

\section*{Competing interests}
The authors declare no competing interests.


\renewcommand{\figurename}{SUPPLEMENTARY FIG.} 
\renewcommand{\tablename}{SUPPLEMENTARY TABLE} 
\renewcommand{\thetable}{\arabic{table}} 
\setcounter{figure}{0}

\section*{Supplementary Note 1: Device Parameters}
Our device parameters are summarized in Supplementary Table~\ref{params}. To determine $\varphi_k$ for the CPW resonators and $\varphi_b$ for the buffer mode, we simulate the capacitive coupling between the CPWs and the buffer and fit the inductance of the ATS junction array that reproduces the measured buffer mode frequency. We leave the ATS junction array and SQUID junctions open in our capacitance simulation. We find the ATS junction array inductance to be $L_b$ = 7.50 nH, corresponding to $E_{L_b}/h =$ 21.8 GHz. This values agrees within 8\% of values computed from room-temperature resistance measurements (using the Ambegaokar-Baratoff formula) of nominally identical junctions that are fabricated near the device. We use an array of 21 junctions to form the ATS junction array. The SQUID junction energy is also inferred from room-temperature measurements. An ATS that was fabricated in parallel with the measured device is pictured in Supplementary Fig.~\ref{fig:JJ_closeup}. 

\begin{figure}[h!]
\includegraphics[scale = 0.9]{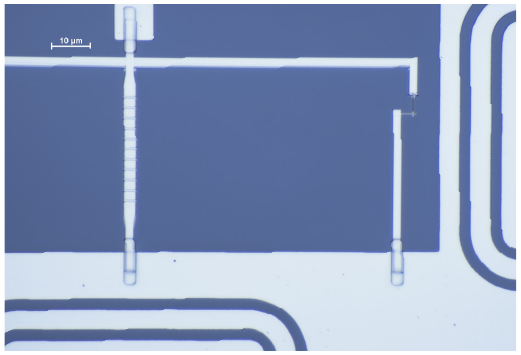}
\caption{\label{fig:JJ_closeup} Optical micrograph of the ATS. This is an image of an identical device that was fabricated in parallel with the measured device, with the exception that the Dolan-style junctions in the ATS superinductor are nominally 0.975~$\mu$m narrower.}
\end{figure}

Given the capacitance matrix, the ATS inductance, and by treating the quarter-wavelength CPW resonators as effective lumped element LC resonators (based on their designed characteristic impedances and lengths), we can diagonalize the circuit to find the eigenfrequencies and ZPFs at the ATS node. We find that our calculated eigenfrequencies agree well with our measured frequencies (within 1\%). The resulting eigenvectors are used to compute $\varphi_b$ and $\varphi_k$.

The buffer mode and the CPW modes are characterized using spectroscopy with a vector network analyzer when the buffer is flux biased at the saddle point. We are able to directly measure the CPW modes due to parasitic capacitances between the CPW resonators and the readout transmission line. The measured values are reported in Supplementary Table \ref{params}. The buffer frequency, extrinsic loss, and intrinsic loss are respectively labeled by $\omega_b$, $\kappa_{b,e}$, $\kappa_{b,i}$. The reported uncertainties are standard deviations from repeatedly measuring the buffer over 95 samples. The CPW frequencies, extrinsic losses (due to parasitic capacitance to the readout line), intrinsic losses, and intrinsic quality factors are respectively labeled by $\omega_k$, $\kappa_{k,e}$, $\kappa_{k,i}$, and $Q_{i,k}$. The reported uncertainties are standard deviations from repeatedly measuring the CPWs over 81 samples, all at single-photon powers.

\begin{table*}

\begin{tabular}{||c| c ||} 
 \hline
 $\omega_b/2\pi$& 5.0073 GHz $\pm$ 600 kHz\\ 
 \hline
 $L_b$ & 7.50 nH \\
 \hline
 $\varphi_b$ & 0.336\\ 
 \hline
 $\kappa_{b,e}/2\pi$ & 3.95 MHz $\pm$ 60 kHz\\ 
 \hline
 $\kappa_{b,i}/2\pi$ & 130 kHz $\pm$ 40 kHz\\ 
 \hline
 $E_J/h$ & 5.28 GHz\\ 
 \hline
\end{tabular}

\vspace*{1 cm}

\begin{tabular}{||c|c|c|c|c||c|} 
 \hline
 $k$ & $\omega_k/2\pi$ [GHz $\pm$ kHz] & $\varphi_k$ & $\kappa_{k,e}/2\pi$ [kHz $\pm$ kHz] & $\kappa_{k,i}/2\pi$ [kHz $\pm$ kHz] & $Q_{i,k} \times 10^3$\\  [1ex] 

 \hline
 1 & 6.904939 $\pm$ 2 & 0.0186 & 37 $\pm$ 1 & 32 $\pm$ 3 & 220 $\pm$ 20 \\ [1ex]
 2 & 6.975562 $\pm$ 3 & 0.0228 & 40 $\pm$ 2 & 29 $\pm$ 2 & 240 $\pm$ 20 \\ [1ex]
 3 & 7.156324 $\pm$ 9 & 0.0210 & 70 $\pm$ 5 & 41 $\pm$ 7 & 170 $\pm$ 30 \\ [1ex]
 4 & 7.247145 $\pm$ 3 & 0.0175 & 57 $\pm$ 1 & 34 $\pm$ 4 & 210 $\pm$ 30 \\ [1ex]
 5 & 7.318975 $\pm$ 4 & 0.0179 & 74 $\pm$ 1 & 39 $\pm$ 4 & 190 $\pm$ 20 \\ [1ex]
 6 & 7.389379 $\pm$ 2 & 0.0186 & 102 $\pm$ 2 & 25 $\pm$ 3 & 300 $\pm$ 40 \\ [1ex]
 7 & 7.460333 $\pm$ 4 & 0.0211 & 132 $\pm$ 4 & 67 $\pm$ 6 & 110 $\pm$ 10 \\ [1ex]
 \hline
\end{tabular}
\caption{Table of device parameters.}
\label{params}

\end{table*}

\begin{figure}[h!]
\includegraphics[scale=0.50]{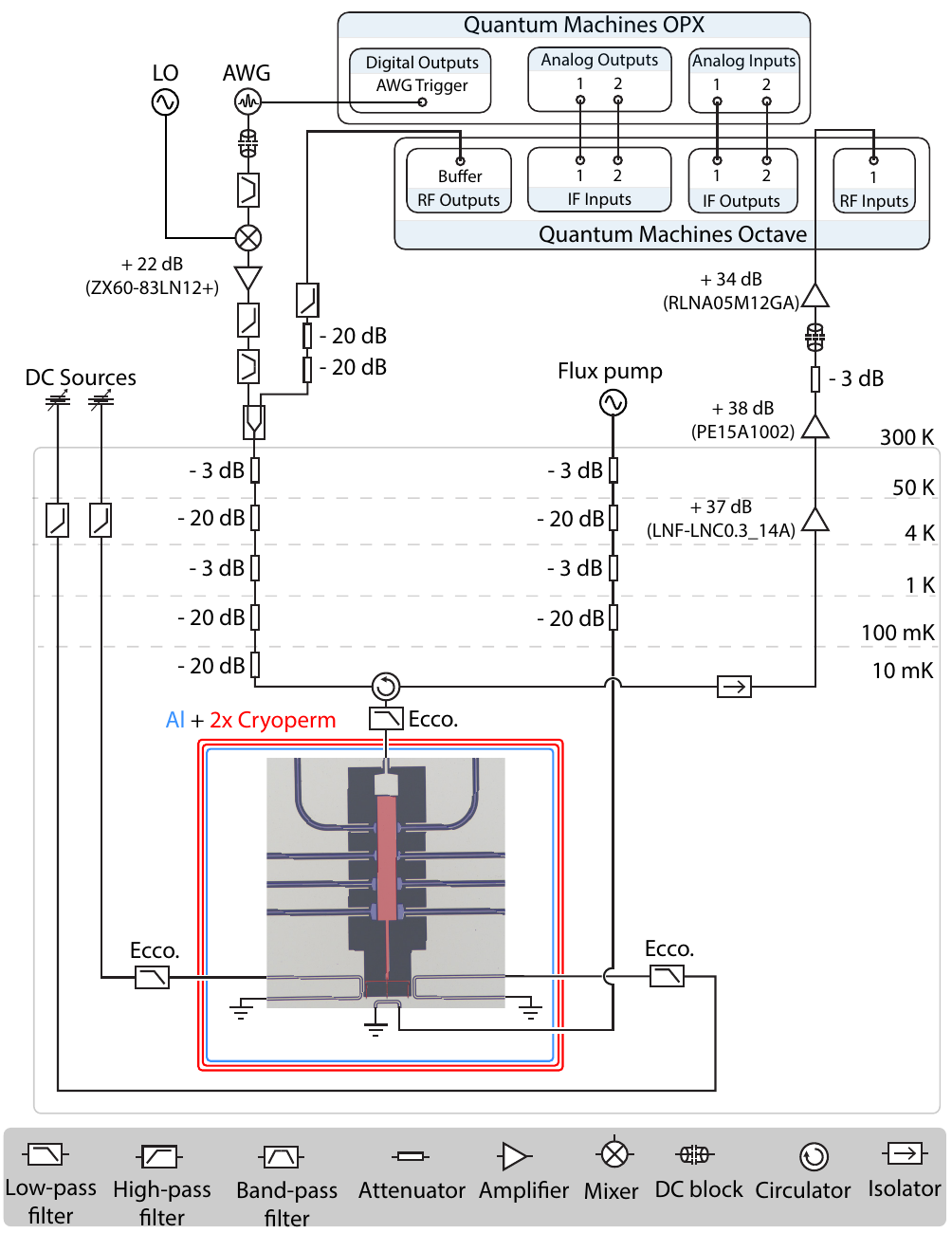}
\caption{\label{fig:wiring} Experimental setup.}
\end{figure}

\begin{figure}[b]
\includegraphics[scale=1]{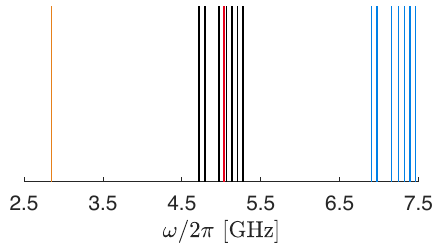}
\caption{\label{fig:sideband} Illustration of relevant frequencies for our experiment. We show the CPW frequencies in blue, the buffer frequency in red, the parametric drive frequencies in black, and the parametric pump frequency in orange.}
\end{figure}

\begin{figure}[b]
\includegraphics[scale=1]{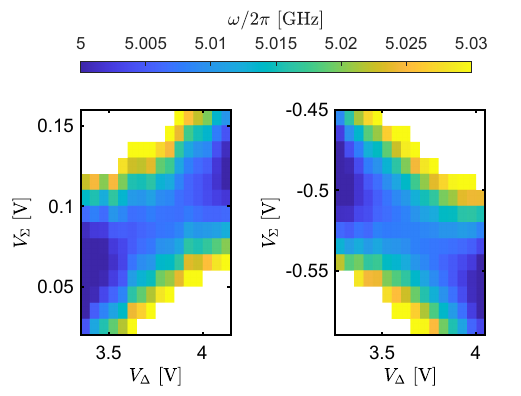}
\caption{\label{fig:saddle} Finding the saddle point. We tune the buffer mode frequency by sweeping our two voltage sources. The point where $\varphi_\Sigma 
= \varphi_\Delta = \pi/2$ is easily identified as a saddle point in a color map of the mode. Here, we show saddle points at two different values of $V_\Sigma$.}
\end{figure}

\section*{Supplementary Note 2: Experimental details and setup}
The experimental setup is shown in Supplementary Fig.~2. Here, we describe the main components of our experiment: (1) parametric driving of the buffer mode, (2) playing and digitizing pulses, (3) DC flux biasing, and (4) flux pumping.

\textit{Parametric driving of the buffer mode:} We use a 5 GS/s AWG to simultaneously play the seven drive tones that parametrically couple the CPW resonators to the buffer mode. Each individual drive tone had a $V_{pp}$ ranging from 3.3 mV to 27 mV depending on the detuning. These tones are amplified after up-conversion to provide strong enough driving on the buffer mode. The output is combined with the output of the QM OPX and Octave. The on-chip power of each individual drive tone approximately ranged from $-99.6$ dBm to $-120.0$ dBm depending on the detuning.

\textit{Playing and digitizing pulses:} We use a QM OPX and Octave to generate Gaussian pulses that we store in the PADL. The pulses are synthesized and played from the Analog Outputs on the OPX, which are then up-converted using mixers and local oscillators (LOs) in the Octave. The Octave conveniently calibrates its internal mixers to remove spurious tones from the LO. The up-converted pulse is played from the RF Outputs on the Octave. Similarly, the Octave down-converts pulses after they interact with our device. Specifically, pulses incident on the RF Inputs are down-converted and played from the IF Outputs, which are then sent to the Analog Inputs on the OPX. For Gaussian pulses with $\langle n \rangle \simeq 1$ and temporal FWHM of 471 ns, the peak power on-chip is approximately $-140.3$ dBm.

\textit{DC flux biasing:} As shown in Fig. 1g, we have two flux lines that are symmetrically placed on either side of the ATS. We use two voltage sources (SRS SIM928) to bias these two flux lines. The flux lines are low-pass filtered at the 4 K stage of our dilution refrigerator (Aivon Therma-24G). 

\textit{Flux pumping:} As shown in Fig. 1g, we have one additional flux line that is placed directly underneath the ATS (where the ATS is grounded). This flux line provides a magnetic flux that symmetrically threads both ATS loops, thereby providing the parametric flux pump $\epsilon_p(t)$. This flux pump is sourced from a power signal generator (Keysight E8257D PSG). The flux pump power on-chip is approximately -57.6 dBm.

In Supplementary Fig.~3 we illustrate all the relevant frequencies for our experiment. We show the CPW frequencies in blue (from Supplementary Table 1) and the buffer frequency in red (from Supplementary Table 2). We also show the parametric drive frequencies in black, which are (in GHz): 4.719403, 4.789553, 4.971818, 5.062160, 5.132469, 5.202372, 5.276428. Finally, we show the pump frequency in orange, which is 2.84654 GHz. 

\section*{Supplementary Note 3: DC Flux Biasing}
We use the two DC flux lines to find the so-called saddle point where $\varphi_\Sigma = \varphi_\Delta =\pi/2$. In Supplementary Fig.~\ref{fig:saddle}, we show the frequency of the buffer mode in the vicinity of the saddle point as we sweep our two voltage supplies. We work in the basis of $V_\Sigma = (V_1 + V_2)/2 \propto \varphi_\Sigma$ and $V_\Delta = (V_1 - V_2)/2 \propto \varphi_\Delta$, where $V_1$ and $V_2$ are the voltages of the individual voltage supplies. 

In reality, slight junction asymmetry between the two SQUID junctions can lead to slight differences in the buffer frequency at different saddle points given by $\varphi_{\Sigma,\Delta} = \pm \pi/2$. In Supplementary Fig.~\ref{fig:saddle}a and \ref{fig:saddle}b, we show two different saddle points taken at two different values of $\varphi_\Sigma$. Fortunately, we find these two saddle point frequencies agree within a linewidth of the buffer mode. We also confirmed this at saddle points taken at two different values of $\varphi_\Delta$. Therefore, we neglect contributions from junction asymmetry in this work.

\section*{Supplementary Note 4: Parametric Delay Line Parameters}
The scattering parameters in Fig.~2a of the main text are fit to the following model $S_{11}$ model:
\begin{equation}
    S_{11}(\omega) = 1-\frac{\kappa_{b,e}}{i(\omega_b-\omega)+\frac{\kappa_b}{2} + i\sum_k \frac{|g_k|^2}{\omega-\omega_k'+i\kappa_k/2}}
\end{equation}
where $\omega_b$ is the buffer frequency, $\kappa_{b,e}$ is the buffer extrinsic loss rate, $\kappa_b = \kappa_{b,e} + \kappa_{b,i}$ is the total loss rate of the buffer (including extrinsic and intrinsic loss), $g_k$ is the parametric coupling between the buffer and the \( k^\text{th} \) CPW, $\omega_k'$ is the frequency of the photons from the \( k^\text{th} \) CPW that are being parametrically coupled to the buffer, and $\kappa_k$ is the total loss rate of the \( k^\text{th} \) CPW. 

The fit parameters are listed in Supplementary Table~\ref{sparams}. In the first column, we list the CPW frequencies from Supplementary Table~\ref{params} to index each row. The reported uncertainties are standard deviations from repeatedly measuring $S_{11}(\omega)$ 500 times overnight and fitting the parameters.

\begin{table*}
\begin{tabular}{||c| c ||} 
 \hline
 $\omega_b/2\pi$ [GHz $\pm$ kHz]& 5.03123 $\pm$  70 \\ 
 \hline
 $\kappa_{b,e}/2\pi$ [MHz $\pm$ kHz] & 3.37 $\pm$ 70 \\
 \hline
 $\kappa_{b,i}/2\pi$ [kHz $\pm$ kHz] & 440 $\pm$ 60\\ 
 \hline
\end{tabular}

\vspace*{1 cm}

\begin{tabular}{||c|c|c|c||} 
 \hline
 $\omega_k/2\pi$ [GHz] & $\omega_k'/2\pi$ [GHz $\pm$ kHz] & $\kappa_{k}/2\pi$ [kHz $\pm$ kHz] & $g_{k}/2\pi$ [kHz $\pm$ kHz] \\  [1ex] 

 \hline
 6.904939 & 5.032140 $\pm$ 1 & 67 $\pm$ 6 & 530 $\pm$ 20 \\ [1ex]
 6.975562 & 5.032625 $\pm$ 4 & 78 $\pm$ 8 & 530 $\pm$ 20 \\ [1ex]
 7.156324 & 5.031112 $\pm$ 9 & 109 $\pm$ 7 & 540 $\pm$ 10 \\ [1ex]
 7.247145 & 5.031607 $\pm$ 4 & 90 $\pm$ 8 & 560 $\pm$ 10 \\ [1ex]
 7.318975 & 5.033132 $\pm$ 5 & 130 $\pm$ 10 & 520 $\pm$ 20 \\ [1ex]
 7.389379 & 5.033617 $\pm$ 4 & 140 $\pm$ 20 & 520 $\pm$ 10 \\ [1ex]
 7.460333 & 5.03058 $\pm$ 10 & 210 $\pm$ 30 & 520 $\pm$ 20 \\ [1ex]
 \hline
\end{tabular}
\caption{Table of fit parameters in $S_{11}(\omega)$.}
\label{sparams}
\end{table*}

\begin{figure}
\includegraphics[scale=0.9]{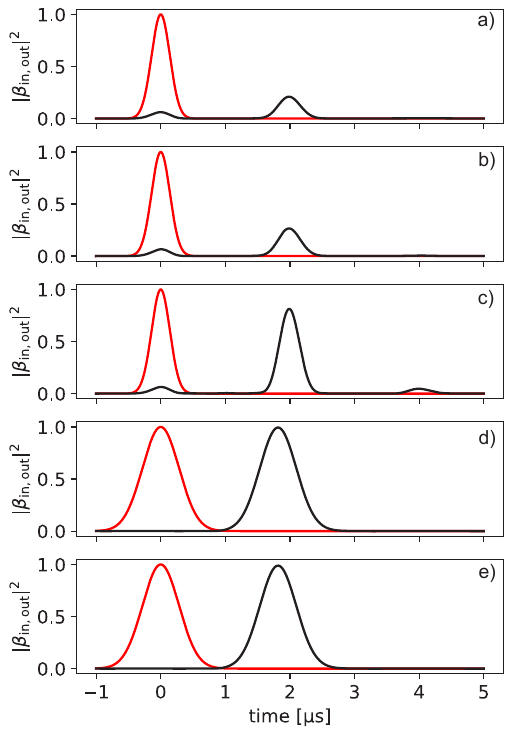}
\caption{\label{fig:Z_mis} Fidelity loss channels. (a) Simulated input (red) and output (black) pulses for the Hamiltonian parameters from our experiment as reported in Supplementary Table~\ref{sparams} and with an input Gaussian pulse (temporal FWHM of 471 ns). In (b) and (c) we run identical simulations but with $\kappa_{b,i} = 0$ and $\kappa_{b,i} = \kappa_k = 0$, respectively. In (d) we not only consider no intrinsic loss, but we also fix $\Delta_k$ to be evenly spaced from $-\kappa_{b,e}/2$ to $+\kappa_{b,e}/2$ over 7 modes and fix $g/2\pi = $ 562.4 kHz. In (e) we consider the same delay line as in (d) but consider nonzero $\kappa_k$ given by state-of-the-art CPW resonators quality factors~\cite{crowley2023disentangling}.}
\end{figure}

\begin{figure}
\includegraphics[scale=0.9]{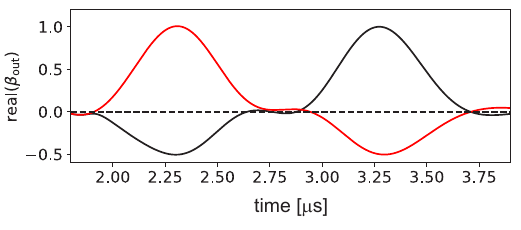}
\caption{\label{fig:swap_sim} Additional pulse-swapping simulation. Simulation of two delayed pulses for a lossless delay line where the parametrically converted CPW photon detunings $\Delta_k$ are not swapped (i.e. $\Delta_k \rightarrow \Delta_k$, shown in red) and when the detunings are swapped (i.e. $\Delta_k \rightarrow -\Delta_k$, shown in black). The two pulses have a relative $\pi$ phase shift and the earlier pulse has an amplitude that is twice as large as the later pulse.}
\end{figure}

\section*{Supplementary Note 5: Dominant sources of infidelity and additional simulations}
Given two temporal modes that are defined by annihilation operators $\hat{A}_1 = \int \mathrm{dt} f(t) \hat{a}_\mathrm{wg}(t)$ and $\hat{A}_2 = \int \mathrm{dt} g(t) \hat{a}_\mathrm{wg}(t)$, we define the fidelity as $F = |\int dt f^*(t)g(t)|^2$~\cite{gheri1998photon}. This is the fidelity associated with two single-photon wavepackets $\hat{A}^\dagger_1|\mathrm{vac}\rangle$ and $\hat{A}^\dagger_2|\mathrm{vac}\rangle$. Even though we are not encoding quantum information in single-photon wavepackets, we can still estimate this fidelity from the detected field, i.e., the complex digitized data collected by the analog-to-digital converter (ADC). We first turn off the pumps and tune the buffer mode off resonance. The device acts as a mirror which reflects the input pulse. Our ADC records the reflected signal and averages the measured field to find $V_\mathrm{in}[t] = I_\mathrm{in}[t] + iQ_\mathrm{in}[t]$, corresponding to the mean field in the reflected pulse. We then operate the PADL in the delay line mode, and record $V_\mathrm{out}[t] = I_\mathrm{out}[t] + iQ_\mathrm{out}[t]$, corresponding to the mean field of delayed pulse. We normalize both of these averaged traces by the same constant such that $\sum_{t} V_\mathrm{in}[t]V_\mathrm{in}[t]^* = 1$. Fidelity for a given delay $\tau$ is then calculated by $F[\tau] = \Big(\sum_{t} V_\mathrm{in}[t-\tau]V_\mathrm{out}[t]^*\Big)^2$. We find that fidelity is maximized for $\tau\simeq T_\mathrm{rt}$, so that $F=\max_\tau(F[\tau])$ which is very nearly $F[T_\mathrm{rt}]$. Note that in the above formulation, since we are using the same normalization constant for both the input and output fields, the fidelity captures the effects of both loss and distortion.

We can simulate this whole process by numerically integrating the equations of motion for $\hat{a}(t)$ and $\hat{b}(t)$ derived from the Hamiltonian in Eq. 4 of the main text. This allows us to estimate the dominant sources of fidelity loss in our delay line. The equations of motion are given by:
\begin{equation}
\begin{split}
    \frac{d}{dt} \hat{a}_k &= -i\Delta_k \hat{a}_k - i g_k^* \hat{b} - \frac{\kappa_{k}}{2}\hat{a}_k -\sqrt{\kappa_{k,i}}\hat{a}_\mathrm{in,i} \\
    \frac{d}{dt} \hat{b} &= -\frac{\kappa_b}{2}\hat{b} - i \sum_k g_k \hat{a}_k - \sqrt{\kappa_{b,e}} \hat{b}_{\text{in}}
\end{split}
\end{equation}
Since these differential equations are linear, we can take the expected values of the fields, and obtain exact mean field equation for $\alpha_k = \langle \hat{a}_k \rangle$ and $\beta = \langle \hat{b} \rangle$. The input and output photon fluxes incident on the buffer mode can be obtained from the input-output boundary condition $\beta_\mathrm{out} = \beta_\mathrm{in} + \sqrt{\kappa_{b,e}}\beta$. 

In Supplementary Fig.~\ref{fig:Z_mis}a, we plot the input photon flux in red and output photon flux in black for different parametric delay line parameters. In the top figure, we simulate the Hamiltonian parameters from our experiment, which are reported in Supplementary Table~\ref{sparams}. We choose $\beta_\mathrm{in}(t)$ to be a Gaussian pulse with temporal FWHM of $471$ ns to match the experiment. The black trace agrees well with the results measured by our ADC in Fig.~2 of the main text, and we calculate a state fidelity of $F = 0.24$. For our fidelity calculation, we normalize the input mode profile ($f(t)$) such that $\int \mathrm{dt} |f(t)|^2 =1$, and we normalize the delayed mode profile ($g(t)$) by the same factor.  In Supplementary Fig.~\ref{fig:Z_mis}b, we set $\kappa_{b,i} = 0$ and find that the fidelity improves to $F = 0.31$. In Supplementary Fig.~\ref{fig:Z_mis}c, we set all intrinsic loss channels to zero i.e. $\kappa_{b,i} = \kappa_k = 0$ and find that the fidelity improves to $F = 0.86$. In Supplementary Fig.~\ref{fig:Z_mis}d, we consider a lossless delay line where the Hamiltonian parameters are chosen such that $\Delta_k$ is evenly spaced from $-\kappa_{b,e}/2$ to $+\kappa_{b,e}/2$ over 7 modes and fix $g/2\pi = $ 562.4 kHz (roughly corresponding to a frequency comb with finesse = 1.5). We also increase the pulse temporal FWHM to 942 ns, which improves the fidelity to $F = 0.996$. Finally, in Supplementary Fig.~\ref{fig:Z_mis}e, we consider the same delay line parameters as in Supplementary Fig.~\ref{fig:Z_mis}d but we add CPW loss again. We consider state-of-the-art CPW quality factors~\cite{crowley2023disentangling} with $Q_i$ of $15e6$ and find that $F = 0.991$.

In Supplementary Fig.~\ref{fig:Z_mis}a, we also observe a small reflected pulse at $t = 0$ in the parametrically delayed (black) trace. This small reflected pulse was present throughout our experiments. This is due to the slight impedance mismatch that arises from our non-ideal phase response. We observe significant reduction in this reflected pulse when we optimize our Hamiltonian parameters and increase our pulse bandwidth, as can be seen in Supplementary Fig.~\ref{fig:Z_mis}d. One challenge we faced in preparing a perfect parametric delay line (i.e. a perfectly linear phase response) is the challenge of rapidly and reliably fitting $S_{11}(\omega)$ while tuning the parametric drive amplitudes and detunings, especially given the large number of parameters to fit. Another source of reflections at $t = 0$ could be inevitable impedance mismatches in our device packaging, most notably from the wirebond connecting our readout line to the PCB.

In Supplementary Fig.~\ref{fig:swap_sim}, we simulate the experiment where we control the detunings of the parametrically converted CPW photons $\Delta_k$ to swap two pulses in time. We consider a lossless delay line (with parameters identical to those in Supplementary Fig.~\ref{fig:Z_mis}d) and we consider two pulses with temporal FWHM = $377$~ns and separation $1000$~ns. In addition to a relative $\pi$ phase shift between the two pulses, we have the first pulse have an amplitude that is twice as large as the second pulse. The red trace in Supplementary Fig.~\ref{fig:swap_sim} plots the pulse when the detunings have not been swapped (i.e. $\Delta_k \rightarrow \Delta_k$), and the black trace plots the pulse when the detunings have been swapped (i.e. $\Delta_k \rightarrow -\Delta_k$). We clearly see that swapping the detunings swaps the pulse in time.

We can use these simulations to approximate the minimum number of resonators needed to perform the simplest possible experiment: delaying a pulse by more than its temporal FWHM. The figure of merit with delay lines is the delay-bandwidth product (DBP). Assuming $\Delta_k$ is evenly spaced from $-\kappa_b/2$ to $+\kappa_b/2$, we have that the DBP $=N-1$ for the PADL, where $N$ is the number of resonators that are being parametrically coupled to the buffer. The narrowest temporal pulse that one could use will have a temporal FWHM $\tau_\text{FWHM} \simeq 2\pi/\kappa_b$. Therefore, we have that $\tau_\text{delay}/\tau_\text{FWHM}\simeq N-1$. To visually differentiate the stored pulse from the delayed pulse, we require $2\tau_\text{FWHM} \lessapprox \tau_\text{delay}$ and thus one would require a bare minimum of 3 resonators.

\begin{figure}[b]
\includegraphics[scale=0.9]{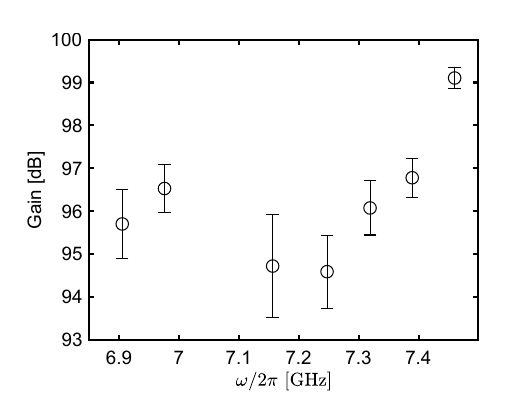}
\caption{\label{fig:gain_cal} Fitted gain at the CPW frequencies from operating them as quantum MPOs near threshold.} 
\end{figure}

\section*{Supplementary Note 6: Gain, Attenuation, and Added noise calibration}
To bound the number of added noise photons caused by strongly driving our device, we first need to carefully calibrate the gain of our measurement apparatus. To do this, we need an in-situ noise source, i.e. some way to relate the number of on-chip quanta to a measurable value at room-temperature. 

In our system, we leverage the three-wave mixing interaction of the ATS to operate our CPW resonators as MPOs. We use the number of resonator photons $n$ near threshold as our in-situ noise source. The smooth increase in $n$ vs. driving field observed in our quantum MPO is similar to the behaviour of a laser near threshold~\cite{bjork1991analysis,bjork1994definition,rice1994photon}.

To operate the \( k^\text{th} \) CPW as an MPO, we pump our device at a frequency $\omega_p = 2\omega_k-\omega_b$ and drive our buffer at a frequency $\omega_d = \omega_b$. To find where the pump and drive frequencies precisely satisfy these energy conservation relations, we sweep them and measure the spectrum of the CPW resonator. Far below threshold, $n$ is only nonzero if the pump and drive frequencies are resonant~\cite{berdou2023one}.

Once we have determined our resonant pump and drive frequencies, we measure the spectrum of the CPW resonator at different drive strengths. We fit the integrated PSD vs. the drive strength to a master equation model with Hamiltonian $\hat H$ and jump operators $\hat L_1$ and $\hat L_2$ given by
\begin{equation}
\begin{split}
    \hat{H} & = i\epsilon_2 \hat{a}^2 + \mathrm{h.c.},\\
    \hat{L}_2 & = \sqrt{\kappa_2} \hat{a}^2, \\ 
    \hat{L}_1 & = \sqrt{\kappa_1} \hat{a},
\end{split}
\end{equation}
where $\epsilon_2$ is the two-photon drive strength, $\kappa_2$ is the two-photon loss rate, and $\kappa_1$ is the single-photon loss rate. An example for the mode at 6.975562 GHz is shown in Fig.~4a of the main text. 

\begin{table}
\vspace{1 cm}
\begin{tabular}{||c|c|c|c||} 
 \hline
 $\omega_k/2\pi$ [GHz] & Gain [dB] & $\kappa_2/\kappa_1$ & $\langle n \rangle $ \\  [1ex] 
 \hline
 6.904939 & 95.70 & 8.406e-4 & 0.143 \\ [1ex]
 6.975562 & 96.53 & 22.09e-4 & 0.092 \\ [1ex]
 7.156324 & 94.72 & 6.323e-4 & 0.109 \\ [1ex]
 7.247145 & 94.59 & 5.523e-4 & 0.143 \\ [1ex]
 7.318975 & 96.07 & 7.165e-4 & 0.083 \\ [1ex]
 7.389379 & 96.78 & 12.73e-4 & 0.065 \\ [1ex]
 7.460333 & 99.11 & 70.65e-4 & 0.021 \\ [1ex]
 \hline
\end{tabular}
\caption{Table of the fitted gain, $\kappa_2/\kappa_1$, and added noise.}
\label{nadded}
\end{table}

Our only fit parameters are: (1) the proportionality constant between $\epsilon_2/\kappa_1$ and the driving field, (2) the proportionality constant between $n$ and the integrated PSD, which is related to the gain, and (3) $\kappa_2/\kappa_1$. By normalizing everything with respect to $\kappa_1$, our fits are robust to pump-induced effects that may worsen $\kappa_1$. In Supplementary Fig.~\ref{fig:gain_cal}, we plot our measured gain (in dB) at the different CPW frequencies (error bars are the standard deviation from repeated measurements and fits). The gain is the proportionality constant between the integrated PSD at room temperature and the power leaking out of the MPO ($P_{\mathrm{MPO},k} = \hbar \omega_k \bar{n} \kappa_{k,e}$). We also report our fitted gains and $\kappa_2/\kappa_1$ in Supplementary Table~\ref{nadded}.
\pagebreak

With the gain calibrated near the CPW frequencies, we can bound the noise added from strongly driving our device. Specifically, we turn on the parametric pump and drives that we use for our delay line experiments while measuring the CPW spectra. An example of such a spectrum is shown in Fig. 4b. In Supplementary Table~\ref{nadded} we report the added noise in each CPW mode and observe it to be much less than 1 photon per mode. In the first column, we list the CPW frequencies from Supplementary Table \ref{params} to index each row.

\end{document}